\documentclass[preprint,...]{revtex4-1}
\usepackage{graphicx}
\usepackage{dcolumn}
\usepackage{bm}

\usepackage[mathletters]{ucs}
\usepackage[utf8x]{inputenc}

\begin{document}

\title{Unfolding the physics of URu$_2$Si$_{2}$ through Si $\rightarrow$ P substitution}

\author{A. Gallagher,$^1$ K. -W. Chen,$^1$ C. M. Moir,$^1$ S. K. Cary,$^3$ F. Kametani,$^2$ N. Kikugawa,$^{1,4}$ D. Graf,$^{1}$ T. E. Albrecht-Schmitt,$^3$ S. C. Riggs,$^1$ A. Shekhter,$^1$ and R. E. Baumbach$^1$}
\affiliation{$^1$National High Magnetic Field Laboritory, Florida State University}
\affiliation{$^2$Applied Superconductivity Center – Florida State University}
\affiliation{$^3$Florida State University – Department of Chemistry And Biochemistry}
\affiliation{$^4$National Institute for Materials Science, 1-2-1 Sengen, Tsukuba, Japan}
\date{\today}

\begin{abstract}
The heavy fermion intermetallic compound URu$_2$Si$_2$ exhibits a ``hidden-order" phase below the temperature of 17.5 K, which supports both anomalous metallic behavior and unconventional superconductivity. While these individual phenomena have been investigated in detail, it remains unclear how they are related to each other and to what extent uranium $f$-electron valence fluctuations influence each one. Here we use ligand site substituted URu$_2$Si$_{2-x}$P$_x$ to establish their evolution under electronic tuning. We find that while hidden order is monotonically suppressed and destroyed for $x$ $\leq$ 0.035, the superconducting strength evolves through a dome that is centered near $x$ $\approx$ 0.01 and terminates near $x$ $\approx$ 0.028. This behavior reveals that hidden order depends strongly on tuning outside of the U$f$-electron shells. It also suggests that while hidden order provides an environment for superconductivity and anomalous metallic behavior, it's fluctuations are not solely responsible for their progression. 
\end{abstract}

\maketitle
Materials that defy straightforward description in terms of either localized or itinerant electron behavior are a longstanding challenge to understanding novel electronic matter.~\cite{herring,mott,keimer,moore_09} The intermetallic URu$_2$Si$_2$ is a classic example, where the itinerant electrons exhibit a giant magnetic anisotropy that is normally characteristic of localized electrons.~\cite{altarawneh_12,chandra} URu$_2$Si$_2$ further displays an unknown broken symmetry state (``hidden-order") and unconventional superconductivity for temperatures below $T_0$ = 17.5 K and $T_c$ = 1.4 K, respectively.~\cite{palstra_85,schlabitz_86,maple_86,mydosh_rev} The development of sophisticated experimental techniques and access to ultra-high-purity single crystal specimens has recently advanced our understanding of the ordered states in this compound. For instance, excitations in the $A_{2g}$ channel have been identified by electronic Raman spectroscopy as a signature of the hidden order,~\cite{blumberg} while elastoresistance,~\cite{riggs} resonant ultrasound,~\cite{yanagisawa,brad} and spectroscopic measurements~\cite{wray} suggest the presence of fluctuations in the $B_{2g}$, $B_{1g}$, and $E_{g}$ channels, respectively. These studies follow high resolution X-ray diffraction,~\cite{tonegawa} torque magnetometry,~\cite{okazaki} and polar Kerr effect measurements,~\cite{schemm} which provide further insight into hidden order. It is also noteworthy that URu$_2$Si$_2$ differs from most other unconventional superconductors, which are typically found near the zero temperature termination point of a line of phase transitions, a ``quantum critical point," where strong fluctuations are believed to be favourable for the superconducting pairing.~\cite{lohneysen_2007_1,pfleiderer_2009_1,loram,walmsley} In contrast, the superconductivity in URu$_2$Si$_2$ is fully contained inside the ordered phase, as revealed by numerous tuning studies.~\cite{mcelfresh_87,amitsuka_99,jeffries_08,hassinger,kanchanavatee_11,kanchanavatee_14,das,amitsuka_88,dalichaouch_90,dalichaouch_89,bauer_05,butch_10} 

To explore the mechanisms of hidden order, anomalous metallic behavior, and superconductivity in this material we synthesized high purity single crystal specimens of chemically substituted URu$_2$Si$_{2-x}$P$_{x}$, where ligand site substitution is a ``gentle" way to tune the electronic state. Recent advances developing a molten metal flux growth technique to produce high quality single crystal specimens of URu$_2$Si$_2$ enabled these experiments, which were previously inaccessible due to metallurgical challenges associated with the high vapor pressure of phosphorous.~\cite{baumbach} We report electrical transport and thermodynamic measurements which reveal that while hidden order is destroyed for $x$ $\leq$ 0.035, the superconducting strength goes through a dome that is centered around $x$ $\approx$ 0.01 and disappears near $x$ $\approx$ 0.028. While the rapid suppression of hidden order indicates the importance of itinerant electrons, the unexpected maximum in superconducting strength may suggest the presence of a critical point that is defined by the termination of a phase boundary other than that of the hidden order. Owing to the relatively small chemical difference between silicon and phosphorous, we propose that in this series the dominant tuning effect is simply to change the chemical potential. Extrinsic factors such as disorder, as well as some intrinsic factors including changes in the unit cell volume or bond angles, spin orbit coupling and ligand hybridization strength play a minor role. Phosphorous substitution further has the advantage that it does not directly affect either the local electron count or the balance of the spin-orbit and Coulomb interactions of the $d$- or $f$-electrons on the Ru and U sites, making it an ideal and long desired tool for unraveling the physics of URu$_2$Si$_2$. 

\section{Results}
\begin{figure}[!tht]
    \begin{center}
        \includegraphics[width=3.5in]{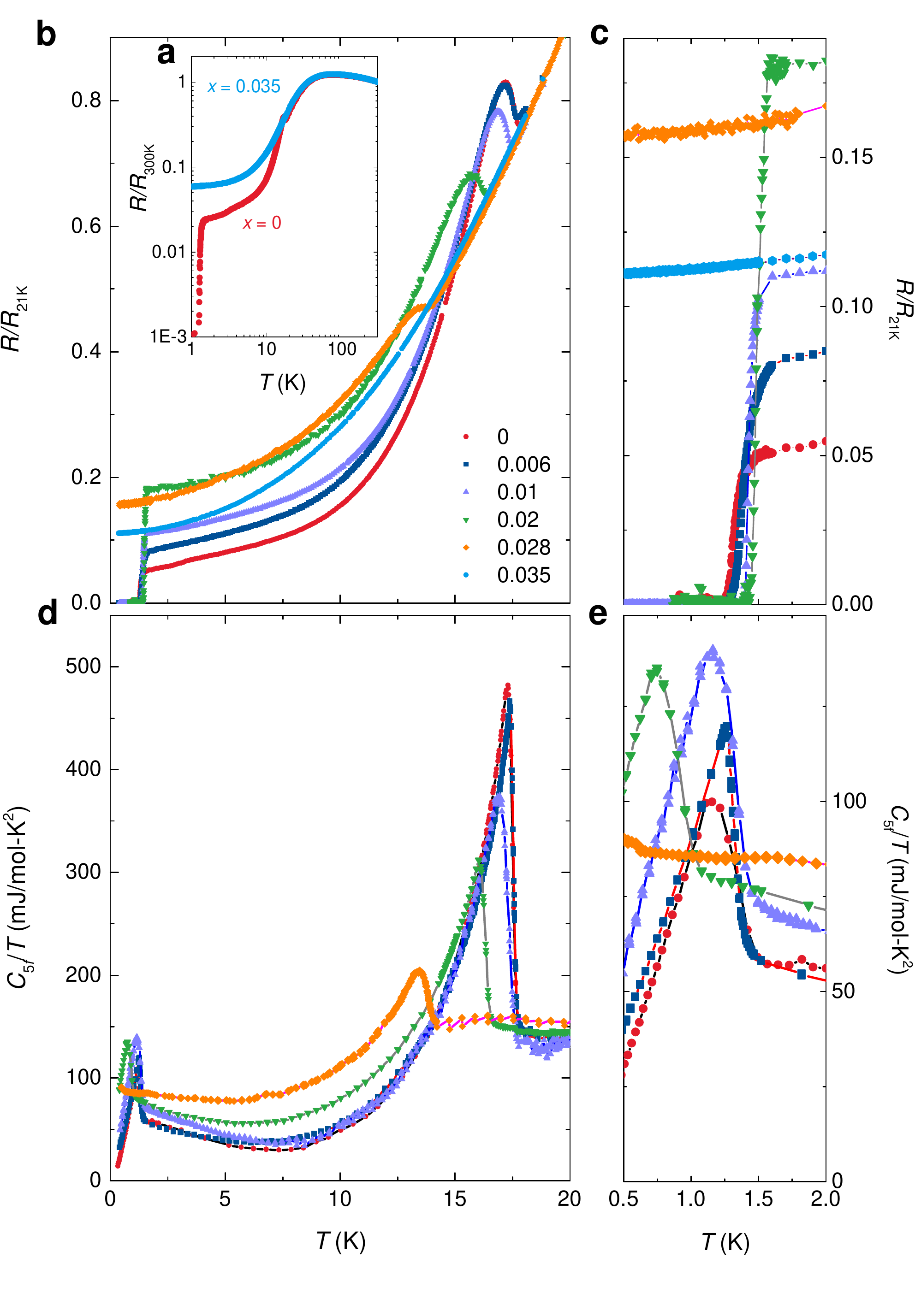}
        \caption{
        (a) Electrical resistance normalized to the value at 300 K $R/R_{300K}$ vs. temperature $T$ for phosphorous concentrations $x$ $=$ 0 and 0.035. (b) Electrical resistance normalized to the value at 21 K $R/R_{21K}$ vs. $T$ for 0 $\leq$ $x$ $\leq$ 0.035. (c) $R/R_{21K}$ vs. $T$ in the low temperature region, emphasizing the superconducting transitions. (d) The 5$f$ contribution to the heat capacity $C_{5f}$ divided by $T$ vs. $T$ for 0 $\leq$ $x$ $\leq$ 0.028. (e) $C_{5f}/T$ vs. $T$ in the low $T$ region, showing the bulk superconducting transitions. See Supplementary for a description of the phonon background subtraction.}
        \label{fig:rho}
    \end{center}
\end{figure}

In Fig.~\ref{fig:rho} we show normalized electrical resistance $R$ and heat capacity $C$ vs. temperature $T$ for several phosphorous concentrations $x$ (see Supplementary for magnetic susceptibility $\chi(T)$). The Kondo lattice behavior of the electrical resistance (i.e., non-monotonic temperature dependence below room temperature with a resistive peak near 80 K) for $T$ $\geq$ $T_0$ is unaffected by phosphorous substitution for $x$ $\leq$ 0.035 (Fig.~\ref{fig:rho}a), suggesting that the strength of the hybridization between the $f$- and conduction electron states does not change much in the range 0 $<$ $x$ $<$ 0.035. At lower temperatures, the hidden order transition temperature $T_0$ and the size of the anomalies in $R$ and $C$ associated with it are monotonically suppressed with increasing $x$ (Figs.~\ref{fig:rho}b,d). There is excellent agreement between the values of $T_0$ as extracted from $R$, $C/T$, and $\chi$, indicating that disorder effects are negligible. We find no evidence for hidden order in $x$ = 0.035 for $T$ $>$ 20 mK, showing that there is a quantum phase transition from hidden order to a paramagnetic correlated electron metal between 0.028 $<$ $x$ $<$ 0.035. However, the data does not preclude the possibility of a broad fluctuation regime around the quantum phase transition. The paramagnetic region subsequently extends up to $x$ $\approx$ 0.25 (a factor of ten larger $x$) where correlated electron antiferromagnetism appears.~\cite{note0} Unlike for other tuning strategies,~\cite{mcelfresh_87,amitsuka_99,jeffries_08,hassinger,kanchanavatee_11,kanchanavatee_14,das,amitsuka_88,dalichaouch_90,dalichaouch_89,bauer_05,butch_10} magnetism is distant from hidden order in this phase diagram. 

The resistive superconducting transition temperature $T_{c,\rho}$ (Fig.~\ref{fig:rho}c) initially increases with $x$ and subsequently vanishes, with no evidence for bulk superconductivity above 20 mK for $x$ $>$ 0.02. While the value of $T_{c,C}$ extracted from heat capacity is in close agreement with $T_{c,\rho}$ for $x$ $\leq$ 0.01, these values separate for $x$ $=$ 0.02 where $T_{c,\rho}$ $>$ $T_{c,C}$. We note that a similar discrepancy between $T_{c,\rho}$ and $T_{c,C}$ is seen for high quality single crystal specimens of the correlated electron superconductor CeIrIn$_5$ and may be an intrinsic feature of the unconventional superconducting state.~\cite{petrovic}  From both $\rho$ and $C/T$, we find that for $x$ $=$ 0.028 there is a transition into the hidden order state near 13.5 K, but no bulk superconductivity down to 20 mK. There is no evidence for superconductivity in $x$ $=$ 0.035 for $T$ $>$ 20 mK. 

\begin{figure}[!tht]
    \begin{center}
        \includegraphics[width=3.5in]{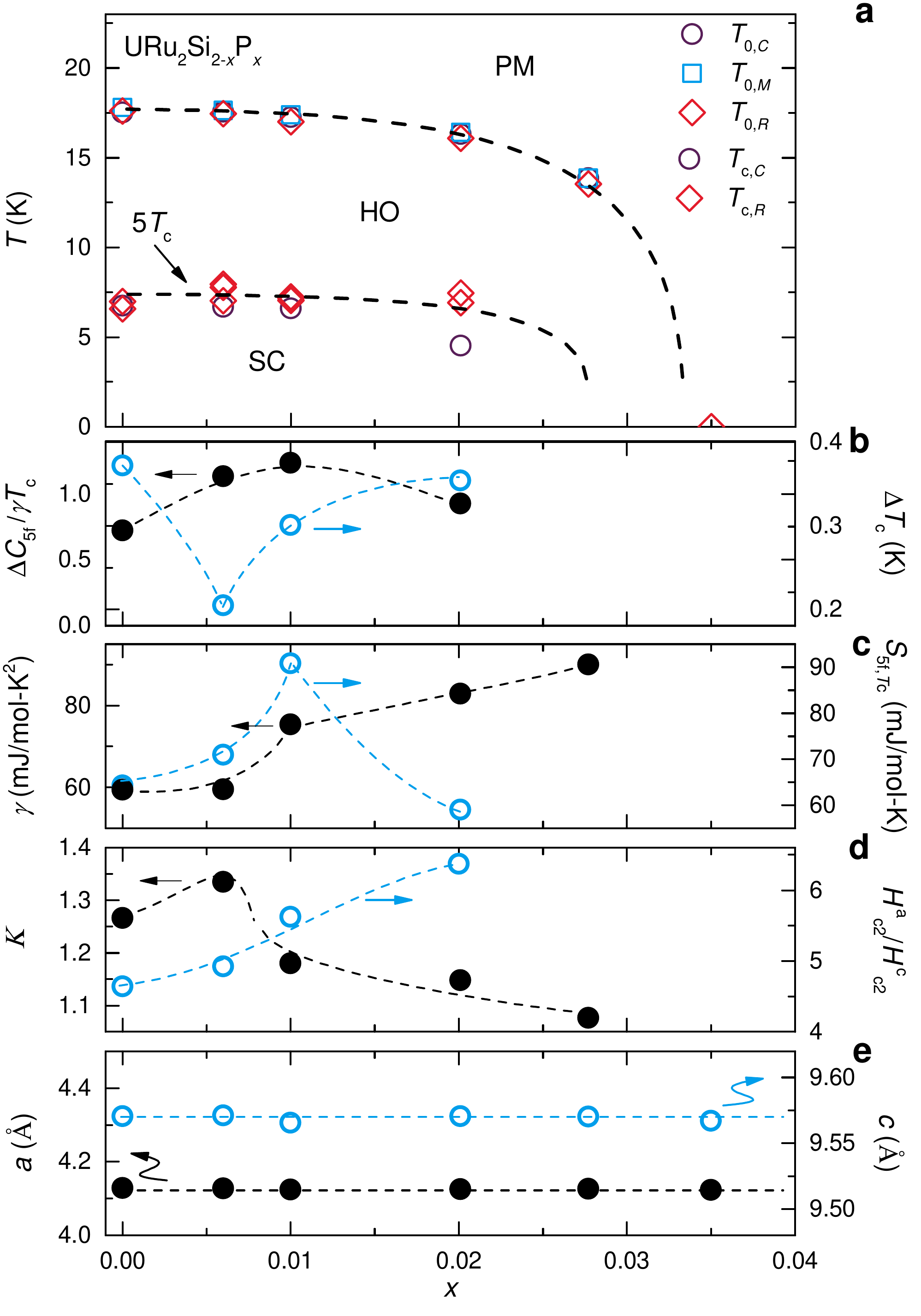}
        \caption{(a) Temperature $T$ vs phosphorous concentration $x$ phase diagram for URu$_2$Si$_{2-x}$P$_x$ constructed from heat capacity (circles), magnetic susceptibility (squares), and electrical resistance (diamonds). The $T-x$ phase boundary $T_0(x)$ separates the paramagnetic heavy electron liquid phase from the hidden order phase. $T_c(x)$ separates the hidden order and superconducting phases. The dotted lines are guides to the eye. (b) Left axis: The size of the discontinuity in the heat capacity divided by the superconducting transition temperature $T_c$ and the electronic coefficient of the heat capacity $\gamma$, $\Delta$$C_{5f}$/$\gamma$$T_c$ vs. $x$. Right axis: The width of the superconducting phase transition $\Delta$$T_c$ vs. $x$. (c) Left axis: The electronic coefficient of the heat capacity $\gamma$ vs. $x$. Right axis: The 5$f$ contribution to the entropy $S_{5f}$ at $T_c$ vs. $x$. (d) Left axis: The value of the Kohler scaled curve at $H/R_0$ $=$ 50 vs. $x$. Right axis: The anisotropy of the upper critical field curves $H^a_{c2}/H^c_{c2}$ vs $x$. (e) The lattice constants, $a(x)$ (left axis) and $c(x)$ (right axis), obtained from single crystal X-ray diffraction measurements.}
        \label{fig:phase}
    \end{center}
\end{figure}

These results are shown in Fig.~\ref{fig:phase}a, where the superconducting region is enclosed by hidden order in the $T-x$ phase space. Over this concentration range, the ground state is mainly tuned by electronic variation, as indicated by the comparably small changes in other intrinsic and extrinsic factors. The lowest residual resistivity ratio ($RRR$ $\approx$ $\rho_{300K}/\rho_{0}$) for the specimens reported here is $RRR$ $=$ 10 (see supplementary), which is comparable to typical values for parent URu$_2$Si$_2$ where $T_0$ and $T_c$ depend weakly on $RRR$ in the range 10 - 500.~\cite{baumbach} The high crystal-chemical quality of these specimens is further highlighted by the observation of quantum oscillations in electrical transport measurements (see Supplementary), indicating that disorder effects are negligible. The unit cell volume and bond angles are also unchanged by phosphorous substitution (Fig.~\ref{fig:phase}e), in contrast to some previous studies.~\cite{kanchanavatee_11,kanchanavatee_14}

Having established the $T-x$ phase diagram (Fig.~\ref{fig:phase}a), we now discuss the region beneath the hidden order phase boundary, where unexpectedly rich behavior occurs. As evidenced by the jump size in $C_{5f}$/$T$ at $T_c$ ($\Delta$$C_{5f}$/$T_c$) (and the transition width), which evolve though a maximum (and a minimum) between 0.006 and 0.01, respectively (Figs.~\ref{fig:rho}e and ~\ref{fig:phase}b), phosphorous substitution non-monotonically enhances the thermodynamic signature of the superconductivity. Here, $C_{5f}$ refers to the heat capacity following subtraction of the nonmagnetic ThRu$_2$Si$_2$ lattice term, as described in the supplementary section. The non-monotonic behavior is reflected in the behavior of $S_{5f,Tc}$($x$) (Fig.~\ref{fig:phase}c), which goes through a maximum near $x$ $=$ 0.01. In contrast, for the hidden order $\Delta$$C$/$T_0$ and $S_{5f,T0}$ are monotonically suppressed with increasing $x$ (see Supplementary). Further evidence for non-monotonic evolution of the superconductivity is provided by the doping evolution of the ratio $\zeta$ $=$ $\Delta$$C_{5f}$/$\gamma$$T_c$($x$), (Fig.~\ref{fig:phase}b) which, for conventional superconductors, is a numeric constant $\zeta_{\rm{BCS}}$ $=$ 1.43. By using $C_{5f}/T$ at $T_c$ for the value of the normal state $\gamma$ we find that $\zeta(x)$ evolves non-monotonically through a maximum value of 1.2 at $x$ $=$ 0.01 (Fig.~\ref{fig:phase}b), where the $x$ $=$ 0 value is near 0.7 as previously reported.~\cite{lohneysen_2007_1,pfleiderer_2009_1} This suggests unconventional superconductivity where the coupling strength may evolve through a maximum.

Magnetoresistance data (Fig.~\ref{fig:RH}) show quantum oscillations, emphasizing the high quality of these specimens (see Supplementary). Similar to the parent compound, the upper critical field $H_{c2}$ (Fig.~\ref{fig:RH}d) is highly anisotropic at all $x$ and follows $H_{c2}(\theta)$ $\propto$ $1/\sqrt{g^2_c\cos^2\theta + g^2_a\sin^2\theta}$ dependence (where $\theta$ is measured from the $c$-axis), suggesting that the upper critical field is Pauli limited.~\cite{altarawneh_12} While there is little $x$ dependence in $g_c(x)$, the $a$-axis $g$-factor $g_a(x)$ significantly decreases before the superconductivity is destroyed near $x$ $\approx$ 0.028 (Fig.~\ref{fig:RH}d). We note that the actual value of the $g$-factor in the $a$-direction may differ significantly from the fitted $g_a(x)$ because of the increased importance of diamagnetic effects as the field rotates into the ab-plane. It remains to be seen whether these trends are consistent with recent theoretical proposals such as ref.~\cite{chandra}.

Magnetoresistance measurements further highlight the non-monotonic evolution with $x$ of the superconductivity and the underlying metallic state. Fig.~\ref{fig:RH}e demonstrates Kohler scaling for $H$ $<$ 9 T applied parallel to the $c$-axis at all dopings,~\cite{kohler} suggesting that the magnetotransport is controlled by the same (temperature dependent) relaxation time as the  zero field resistivity. At each composition in the range 0 $<$ $x$ $<$ 0.028 the normalized magnetoresistance is described by a distinct $f_x(h)$ (where $h$ $=$ $H/\rho(0,T)$), which itself evolves with doping. Notably, the function $f_x(h)$ evolves non-monotonically with $x$ with a maximum near $x$ $=$ 0.006 (Fig.~\ref{fig:phase}d). The maximum in the value of $f_x(h)$ nearly coincides with the maximum in the thermodynamic signatures of the superconductivity inside the hidden order phase.

\section{Discussion}
Although much of the recent excitement surrounding URu$_2$Si$_2$ has focused on the uranium electronic structure and the symmetry of the hidden order phase, a more fundamental question is the degree to which the $f$-electrons can be treated as being localized and the role of quantum fluctuations. The continuity of experimental information extracted from well-developed applied pressure ($P$) and chemical substitution ($x$) series has proven essential to disentangle such effects in other correlated systems including high temperature superconducting cuprates, pnictides, and heavy fermion compounds. To some extent, URu$_2$Si$_2$ has also benefited from such studies. For example, pressure drives a first order phase transition from hidden order into antiferromagnetism near $P_c$ $=$ 0.5 GPa, with a simultaneous evolution of the Fermi surface,~\cite{mcelfresh_87,amitsuka_99,jeffries_08,hassinger} but the resulting insight is limited by the small number of pressure-cell compatible experimental probes. Ruthenium site substitution with Fe and Os produces $T-x$ phase diagrams that closely resemble the $T-P$ phase diagram,~\cite{kanchanavatee_11,kanchanavatee_14,das} but the information gained from these series is constrained by strong disorder. Moreover, ruthenium site substitution is particularly disruptive, as evidenced by the rather different phase diagrams resulting from Rh and Re substitution studies where the hidden order and superconductivity are rapidly destroyed.~\cite{amitsuka_88,dalichaouch_90,dalichaouch_89,bauer_05,butch_10} To understand the complex interplay between different phenomena in this compound a more ``gentle" tuning scheme has long been desired, which could provide access to the physics of URu$_2$Si$_2$ in clean single crystals at ambient pressure. In this context, ligand site substitution in URu$_2$Si$_2$ is an obvious target for investigation.

While in many theoretical scenarios for hidden order in URu$_2$Si$_2$ the U-5$f$ electrons are treated as having mostly fixed valence in a particular atomic crystal field state,~\cite{chandra,blumberg} it is widely believed that they actually have a dual character: i.e., the dynamic nature of the U-5$f$ valence electrons allows for fluctuations between different configurations. However, measurements of the pure compound so far give no insight into the role of these fluctuations in producing hidden order and superconductivity. The rapid changes in the hidden order and superconductivity in our measurements confirm the importance of the itinerant electrons. This is further supported by the observation of weak observation of Kondo lattice physics \cite{hewson} (which tracks the hybridization strength between $f$- and conduction electrons) and strong evolution in the $g$-factor anisotropy (which is a marker for local moment character). Together these results point towards this series as a platform for unraveling the relationship between local and itinerant behavior in URu$_2$Si$_2$.

The stark contrast in the evolution of the hidden order and superconductivity in URu$_2$Si$_{2-x}$P$_x$ (monotonic vs. non-monotonic) further suggests that hidden order, although necessary, is not directly responsible for the superconducting pairing. Instead, the observation of a superconducting dome completely contained inside the hidden order region may indicate the presence of an independent collapsing phase boundary within the hidden order state, as is ubiquitous in other unconventional superconductors. This scenario is reinforced by the observed non-monotonic evolution of the normal state electrical transport, which is also common in correlated electron systems~\cite{lohneysen_2007_1,pfleiderer_2009_1,loram,walmsley} where the strongest deviation from Fermi liquid behavior is seen near the critical point. Alternatively, the independent evolution of hidden order and superconductivity may suggest several competing order parameters in the hidden order phase, as evidenced by electronic Raman ($A_{2g}$),~\cite{blumberg} elastoresistance ($B_{2g}$),~\cite{riggs} resonant ultrasound ($B_{1g}$),~\cite{yanagisawa,brad} and spectroscopic measurements ($E_{g}$).~\cite{wray} These studies should be extended into ligand site substituted URu$_2$Si$_2$. 

Finally, the existing theoretical landscape focuses on $f$-electron physics with no guidance regarding the specificity of the transition metal ion. It is especially puzzling that hidden order and superconductivity are only observed in the U-Ru duo. Examination of silicon site substituted transition metal analogues (U$T_2$Si$_{2-x}$P$_x$, $T$ = transition metal), which can now be synthesized using molten metal flux growth,~\cite{note2} may be particularly illuminating in addressing the universality of hidden order and superconductivity in this fascinating uranium compound. 

\section{Methods}

\textbf{Single crystal synthesis using molten indium flux} Single crystals of URu$_2$Si$_{2-x}$P$_x$ were grown from elements with purities $>99.9$\% in a molten In flux, as previously reported.\cite{baumbach} The reaction ampoules were prepared by loading the elements into a 5 cm$^3$ tantalum crucible in the ratio 1(U):2(Ru):2(Si):22(In). The crucible was then loaded into an alumina tube spanning the bore of a high temperature horizontal tube furnace. Argon gas was passed through the tube and a zirconium getter was placed in a pot before the tantalum crucible in order to purify the argon at high temperatures. The crucible was heated to 500 $^{\rm{o}}$C at 50 $^{\rm{o}}$C/hr, dwelled for 5 hours, heated to 600 $^{\rm{o}}$C at 50 $^{\rm{o}}$C/hr, dwelled for 5 hours, and heated to 1450 $^{\rm{o}}$C at 70 $^{\rm{o}}$C/hr. The dwells at intermediate temperature are intended to allow the phosphorous to completely dissolve into the indium flux without producing a dangerous high vapor pressure. The crucible was then cycled between 1450 - 1400 $^{\rm{o}}$C at 100 $^{\rm{o}}$C/hr ten times. Finally, the furnace was turned off and quickly cooled to room temperature. The indium flux was subsequently removed using hydrochloric acid, to which the URu$_2$Si$_{2-x}$P$_x$ crystals are insensitive. This technique produced single crystal platelets similar to the ones previously reported.

\textbf{Bulk thermodynamic and electrical transport measurements} Heat capacity measurements were performed for mosaics of single crystals using the He3 option in a Quantum Design Physical Properties Measurement System for temperatures 400 mK $<$ $T$ $<$ 20 K. Magnetization $M(T,H)$ measurements were carried out for mosaics of single crystals for temperatures $T$ $=$ 1.8 - 350 K under an applied magnetic field of $H$ $=$ 5 kOe applied parallel to the $c$-axis using a Quantum Design Magnetic Property Measurement System. Magnetic susceptibility $\chi$ is defined as the ratio $M/H$. Zero magnetic field electrical resistance $R$ was measured using the He3 option in Quantum Design Physical Properties Measurement System for temperatures 400 mK $<$ $T$ $<$ 300 K. Several individual crystals were measured for each concentration, which revealed a high degree of batch uniformity. The angular dependence of the superconducting upper critical field was measured using the superconducting magnet (SCM-1) dilution refrigerator system at the National High Magnetic Field Laboratory for $H$ $<$ 18 T and $T$ $=$ 20 mK. Additional magnetoresistance measurements were performed at the National High Magnetic Field Laboratory, Tallahassee, up to magnetic fields of 35 tesla and at $T$ $=$ 50 mK.

\section{Acknowledgements}
This work was performed at the National High Magnetic Field Laboratory (NHMFL), which is supported by National Science Foundation Cooperative Agreement No. DMR-1157490, the State of Florida and the DOE." A portion of this work was supported by the  NHMFL User Collaboration Grant Program (UCGP). TAS and SC acknowledge support from the U.S. Department of Energy, Office of Science, Office of Basic Energy Sciences, Heavy Elements Chemistry Program, under Award Number DE-FG02-13ER16414.

\begin{figure}[!tht]
    \begin{center}
        \includegraphics[width=3.5in]{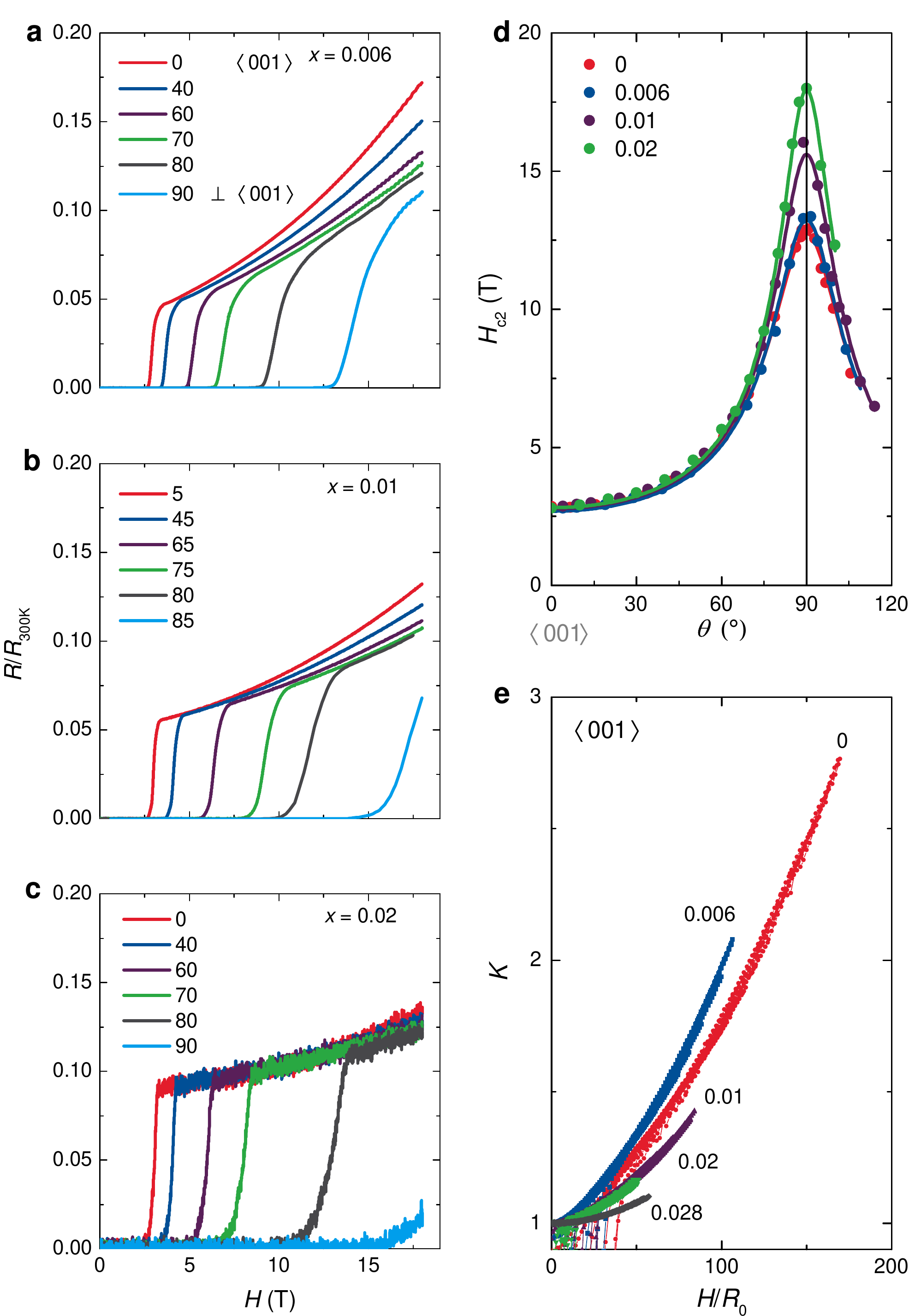}
        \caption{(a) The $x$ $=$ 0.006 electrical resistance normalized to the room temperature value $R/R_{300K}$ vs. magnetic field $H$ for several different angles $\theta$. The data were collected at the temperature $T$ $=$ 20 mK. The electrical current was applied in the $ab$-plane and $\theta$ $=$ 0 is the configuration where $H$ is parallel ($\parallel$) the crystallographic $c$-axis. (b) $R/R_{300K}$ vs. $H$ for $x$ $=$ 0.01 at $T$ $=$ 20 mK for select $\theta$. (c) $R/R_{300K}$ vs. $H$ at $T$ $=$ 20 mK for $x$ $=$ 0.02 for select $\theta$. (d) The upper critical field $H_{c2}$, defined as the extrapolated zero resistance intercept, for $T$ $=$ 20 mK for 0 $\leq$ $x$ $\leq$ 0.02. Data for $x$ $=$ 0 is from ref.~\cite{ohkuni} (e) The Kohler scaled electrical resistivity $K = \frac{\rho(H,T)}{\rho(0,T)}$ vs. the reduced field $H/R_0$ for 0 $\leq$ $x$ $\leq$ 0.028.}
        \label{fig:RH}
    \end{center}
\end{figure}


\begin{thebibliography}{99}
\bibitem{note0}
Results for the larger $x$ URu$_2$Si$_{2-x}$P$_x$ substitution series will be reported separately.
\bibitem{note2}
We have already successfully synthesized several transition metal analogues U$T_2$Si$_2$ ($T$ $=$ Mn, Fe, Ir and Pt) using the same flux growth recipe.
\end{thebibliography}
\end{document}